\documentclass[allclo]{FBSart}
\usepackage{amsfonts}
\usepackage{amssymb}
\usepackage{epsfig}

\title{Trajectory of virtual, bound and resonant Efimov states}
\author{M. T. Yamashita$^1$\thanks{\textit{E-mail address:}
yamashita@itapeva.unesp.br}, T. Frederico$^2$ and Lauro Tomio$^3$}
\institute{$^1$Campus Experimental de Itapeva, UNESP, 18409-010 Itapeva, SP, Brazil\\
$^2$Departamento de F\'\i sica, ITA, CTA, 12228-900 S\~ao Jos\'e dos Campos, Brazil\\
$^3$Instituto de F\'\i sica Te\'orica, UNESP, 01405-900, S\~{a}o Paulo, Brazil}

\runningauthor{M.\,T.\,Yamashita}
\runningtitle{Trajectory of virtual, bound and resonant Efimov states}
\begin{document}

\maketitle
\begin{abstract}
The pole trajectory of Efimov states for a three-body $\alpha\alpha\beta$ system
with $\alpha\alpha$ unbound and $\alpha\beta$ bound is calculated using a zero-range
Dirac-$\delta$ potential.
It is showed that a three-body bound state turns into a virtual one  by increasing
the $\alpha\beta$ binding energy.
This result is consistent with previous results for three equal mass particles. The
present approach considers
the $n-n-^{18}C$ halo nucleus. However, the results have good perspective to be tested
and applied in ultracold atomic systems, where one can realize such three-body configuration
with tunable two-body interaction.
\end{abstract}

The nonintuitive appearance of an infinite number of three-body bound states,
called Efimov states, when the two-body energy tends to zero is being nowadays
largely studied in nuclear and atomic systems. Recently the first indirect experimental
evidence of these states was found for cesium atoms in an ultracold trap~\cite{grimm}.
The trajectory of Efimov states as a function of the two-body energy (bound or virtual)
considering three equal-mass particles interacting by a zero-range potential follows the
route virtual-bound-resonance \cite{virtual} for a large two-body scattering length
varying from positive to negative values passing through the infinite (this corresponds
to a change from a bound to a virtual state in the two-boson system).

\begin{figure}[thb!]
\centerline{\epsfig{figure=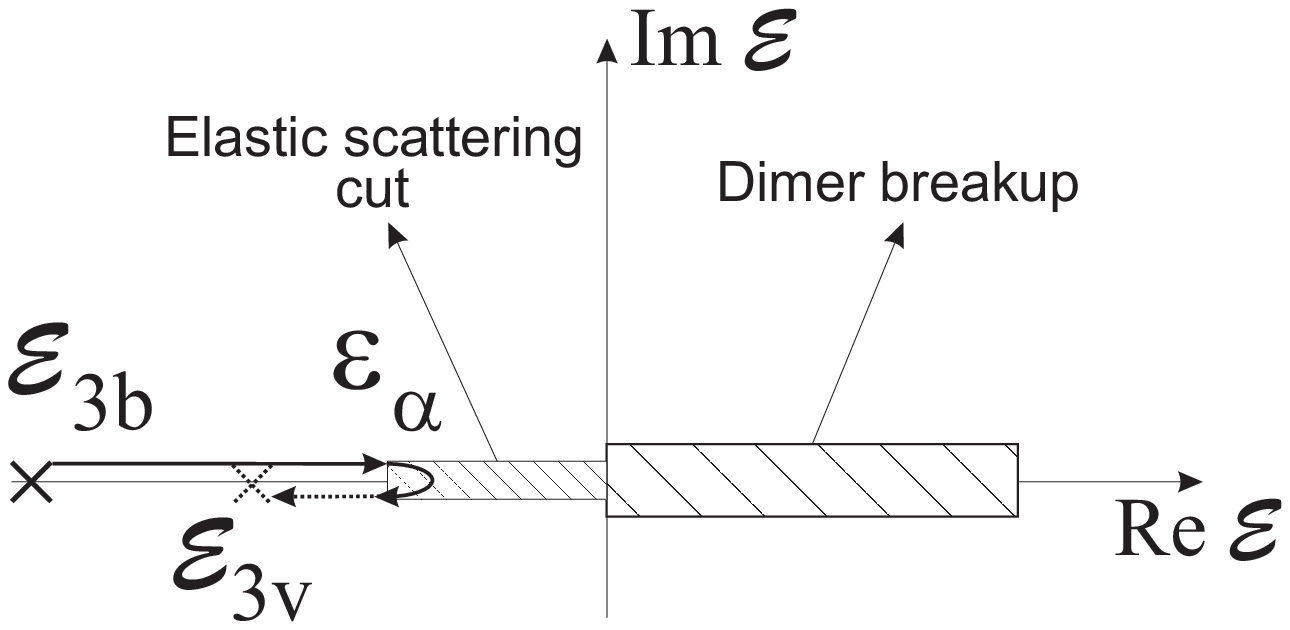,width=6.5cm}
\epsfig{figure=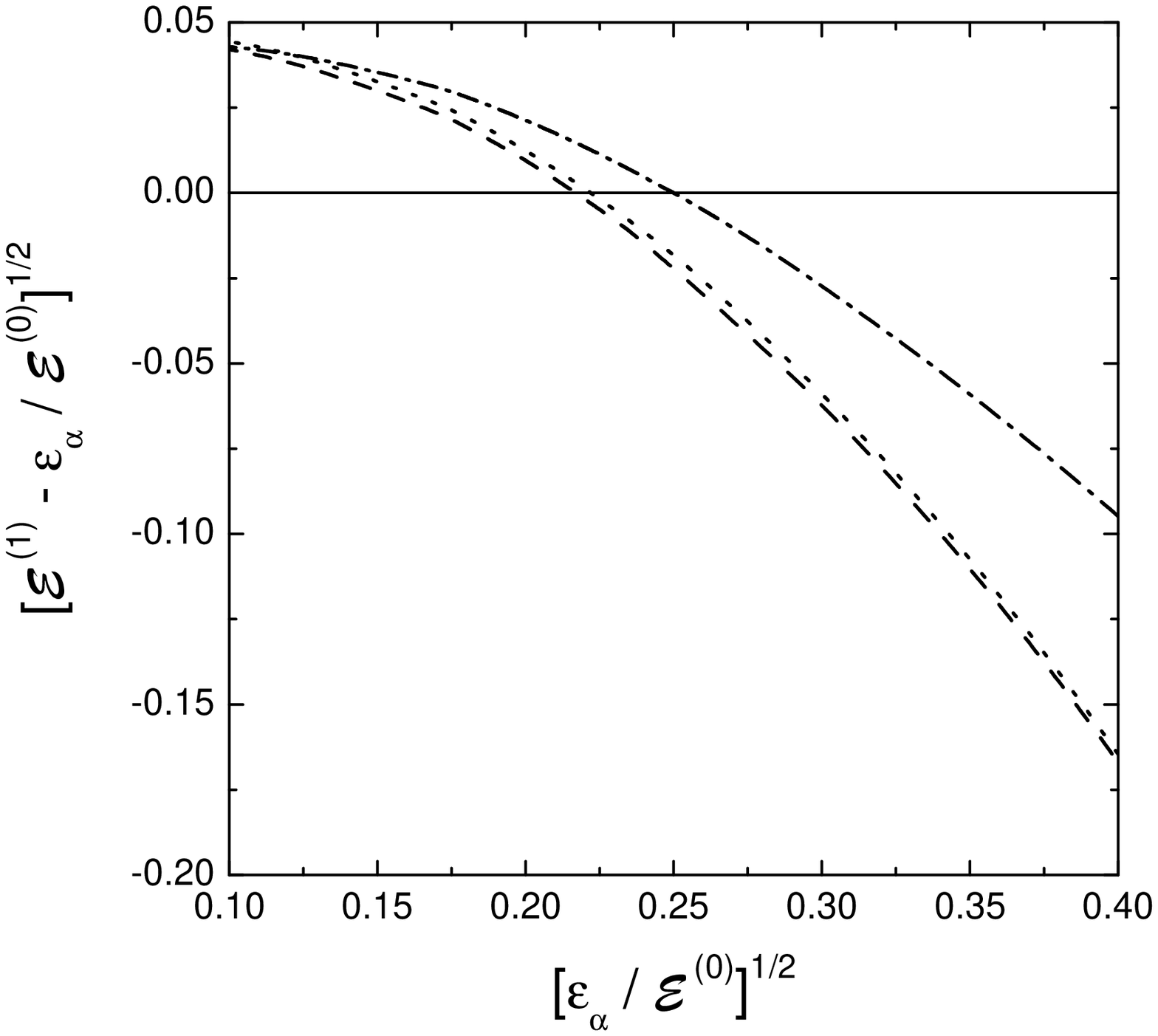,width=6.5cm}}
\caption{{\it Left side}: Three-body energy diagram. The analytic extension of bound state equations
to the second Riemann sheet is made through the elastic scattering cut. The arrow shows the
trajectory of a three-body bound state energy, ${\cal E}_{3b}$, localized in the first Riemann sheet
to a virtual energy, ${\cal E}_{3v}$,
in the second Riemann sheet. {\it Right side}: Bound and virtual three-body energies. The solid
line marks the transition of a bound (positive) to a virtual state (negative)
. The lines dot-dash, dash and dot are, respectively, results for $A=$ 1, 18 and 100.
The superscripts $(0)$ and $(1)$ indicate the ground and first excited state.}
\label{virtual}
\end{figure}

In this communication we present results showing that an excited energy pole for the
$\alpha\alpha\beta$ system, with the subsystems $\alpha\beta$ bound and $\alpha\alpha$
unbound, moves in the complex energy plane from a bound to a virtual state, passing through 
the $\alpha-(\alpha\beta)$ elastic scattering cut, as shown in the diagram given on
the left-side of fig. (\ref{virtual}).

The coupled equations for bound and virtual three-body states (detailed
in Ref.~\cite{plb}), in units of $\hbar=m_\alpha=1$,  can be summarized
in a single-channel equation with $I=b$ for a bound state and
$I=v$ for a virtual state:

\begin{eqnarray}
\label{hn2}
h_\alpha^\ell(q;{\cal E}_{3I})&=&
2{\kappa_v} h_\alpha^\ell(-{\rm i}\kappa_v;{\cal E}_{3v})
{\cal V}^\ell(q,-{\rm i}\kappa_v;{\cal E}_{3v})\delta_{I,v}
\\ &+& \nonumber
\frac{2}{\pi}\int_0^\infty dk k^2
{\cal V}^\ell(q,k;{\cal E}_{3I})
\frac{h_\alpha^\ell(k;{\cal E}_{3I})}{k^2+\kappa_I^2}
\end{eqnarray}

\begin{eqnarray}
{\cal V}^\ell(q,k;{\cal E}_{3I})&\equiv&
\pi\frac{(A+1)}{A+2}{\bar\tau_{\alpha}}(q;{\cal E}_{3I}) \\
&\times& \nonumber
\left[K_2^\ell(q,k;{\cal E}_{3I})+
\int_0^\infty dk'k'^2
K_1^\ell(q,k';{\cal E}_{3I})\tau_{\beta}(k';{\cal E}_{3I})
K_1^\ell(k,k';{\cal E}_{3I})
\right], \label{V}
\end{eqnarray}
with
\begin{eqnarray}
\tau_{\beta}(q;{\cal E}) &\equiv& \frac{-2}{\pi}\left[
\sqrt{|\varepsilon_{\beta}|}+\sqrt{ \frac{A+2} {4A}q^2-{\cal E}}
\right]^{-1}, \label{taunn} \\
{\bar\tau_{\alpha}}(q;{\cal E})&\equiv&
\frac{-1}{\pi}\left(\frac{A+1}{2A}\right)^{\frac 32}
\left(\sqrt{|\varepsilon_{\alpha}}|+\sqrt{
  \frac{(A+2)q^2}{2(A+1)}-{\cal E}}\right).
\label{tauncbar}
\end{eqnarray}
The first term on the right-hand-side of (\ref{hn2}), with
a Kronecker $\delta_{I,v}$, is non-zero only for virtual states.
$A$ is the mass of the particle $\beta$. In the above equations we are using
the odd-man-out notation. The absolute value of the momentum of the spectator
particle with respect to the center-of-mass of the other two particles
is given by $q\equiv|\vec{q}|$; with $k\equiv|\vec{k}|$ being the absolute
value of the relative momentum of these two particles. ${\cal E}$ refers to
a three-body energy, where the indexes $3b$ or $3v$ distinguish between a
bound or virtual state. $\epsilon_\alpha$ is the $\alpha\beta$ biding energy
and $\epsilon_\beta$ is the $\alpha\alpha$ virtual energy. For 
the virtual state energy we have $-{\rm i}\kappa_v=\sqrt{\frac{2(A+1)}{A+2}
\left({\cal E}_{3v}-\varepsilon_\alpha\right)}$.

In order to study the trajectory of Efimov states, for the three-body system given by
two halo neutrons and the $^{18}$C core, we fixed the three-body ground state
energy ($E_{3b}^{(0)}=\hbar^2{\cal E}_{3b}^{(0)}/m_\alpha$ = -3.5 MeV) and the $\alpha\alpha$
two-body energy ($E_{\beta}=\hbar^2\epsilon_\beta/m_\alpha$ = -143 keV, in this case the
$\alpha\alpha$ system is unbound) and vary only the $\alpha\beta$ energy. 
Our present calculations were motivated by the study performed in Ref.~\cite{mazumdar} 
where they have found a different trajectory for the Efimov state in the case of $^{20}$C 
(see also \cite{plb,comment}).

The above technique is
convenient because in the Efimov limit (for a given $A$) the value where a three-body bound state
disappears depends only on the ratio of the two-body energies with the three-body ground state
energy (see pages 327-329 of Ref. \cite{braaten}). In this limit, the results should also be
independent of the potential. Moreover, the Efimov effect is strictly valid when the scattering
lengths are much larger than the effective range. So, to study this effect and its consequences,
it is appropriate to use potential models in the zero-range limit. The numerical solutions of eq.
(\ref{hn2}) are plotted on the right side of fig. (\ref{virtual}) in a form of a universal scaling
function.

In accordance with previous calculations \cite{virtual,amado} for three equal mass particles we can see
that, when at least one two-body subsystem is bound a three-body bound state enters in the second energy
sheet becoming a virtual state. We have also checked that for a Borromean system (all two-body
subsystems are unbound) formed by two equal-mass particles and a different one, a three-body bound state
turns into a resonance following the same behavior of three equal-mass particles. 

\begin{acknowledge}
We would like to thank FAPESP and CNPq for partial support.
\end{acknowledge}

\end{document}